\documentclass{aa}  

\usepackage{graphicx}
\usepackage{txfonts}
\usepackage{booktabs}
\usepackage{silence}
\begin{document}

   \title{A search for radio emission from exoplanets around evolved stars}


   \author{E. O'Gorman
          \inst{1,2}\fnmsep\thanks{ogorman@cp.dias.ie},
          C. P. Coughlan\inst{1},
          W. Vlemmings\inst{2},
          E. Varenius\inst{2},
          S. Sirothia\inst{3,4,5},
          T. P. Ray\inst{1},
          \and
          H. Olofsson\inst{2}
          }

   \institute{Dublin Institute for Advanced Studies, 31 Fitzwilliam Place, Dublin 2, Ireland
   \and
         Department of Space, Earth and Environment, Chalmers University of Technology, Onsala Space Observatory, 43992 Onsala, Sweden
         \and
         National Centre for Radio Astrophysics, TIFR, Post Bag 3, Ganeshkhind, Pune 411007, India
         \and
         Square Kilometre Array South Africa, 3rd Floor, The Park, Park Road, 7405, Pinelands, South Africa
         \and
         Department of Physics and Electronics, Rhodes University, PO Box 94, Grahamstown 6140, South Africa
             }

   \date{Received September 18, 2017; accepted ?}

 
  \abstract 
   {The majority of searches for radio emission from exoplanets have to date focused on short period planets, i.e., the so-called hot Jupiter type planets.  However, these planets are likely to be tidally locked to their host stars and may not generate sufficiently strong magnetic fields to emit electron cyclotron maser emission at the low frequencies used in observations (typically $\geqslant\,150\,$MHz). In comparison, the large mass-loss rates of evolved stars could enable exoplanets at larger orbital distances to emit detectable radio emission. Here, we first show that the large ionized mass-loss rates of certain evolved stars relative to the solar value could make them detectable with the Low Frequency Array (LOFAR) at 150\,MHz ($\lambda = 2\,$m), provided they have surface magnetic field strengths $> 50\,$G. We then report radio observations of three long period ($>1\,$au) planets that orbit the evolved stars $\beta$ Gem, $\iota$ Dra, and $\beta$ UMi using LOFAR at 150\,MHz. We do not detect radio emission from any system but place tight $3\sigma$ upper limits of 0.98, 0.87, and $0.57\,$mJy on the flux density at 150\,MHz for $\beta$ Gem, $\iota$ Dra, and $\beta$ UMi, respectively. Despite our non-detections these stringent upper limits highlight the potential of LOFAR as a tool to search for exoplanetary radio emission at meter wavelengths.
 
}

               
   \titlerunning{A search for radio emission from exoplanets around evolved stars}
   \authorrunning{E. O'Gorman et al.}
   \maketitle
%

\section{Introduction}

The magnetic planets of our solar system have long been known to emit intense coherent radio emission at frequencies below $40\,$MHz \citep{burke_1955}. The emission is predominately due to the interaction between the planet's magnetosphere and the solar wind, although for Jupiter, it is due to magnetosphere-ionosphere coupling, which is independent of the solar wind but dependent on the planetary rotation \citep[e.g.,][]{cowley_2001}. Another intense radio component comes from the Io-Jupiter electrodynamic interaction \citep{marques_2017}. All of these interactions produce energetic electrons that propagate along magnetic field lines into auroral regions of the planet, where electron cyclotron maser (ECM) emission is produced \citep{wu_1979,treumann_2006}. Essentially, the two requirements for ECM emission to occur are that the local plasma frequency must be much less than the gyrofrequency and an unstable keV electron distribution must exist. Both of these requirements are satisfied in high magnetic latitude field lines in both hemispheres of the magnetized solar system planets from just above their surface out to a few planetary radii \citep{zarka_2001}. In the case of Jupiter, the produced decametric emission can be as intense as solar radio bursts in terms of absolute flux density. Consequently, it has long been speculated that exoplanetary radio emission could be detectable \citep[e.g.,][]{yantis_1977} which would not only provide a novel means to directly detect exoplanets but would also allow measurements of exoplanetary magnetic field strengths and rotational periods. 

A crude estimate of the possible radio flux density of an exoplanet can be made by simply scaling the known values of Jupiter to nearby stellar distances. Jupiter is the strongest radio emitter of the solar system planets and at 15\,MHz has a peak flux density of $S_{\nu} \sim 10^{10}\,$mJy \citep{zarka_1992}. If Jupiter were at the distance of a relatively nearby star at 10\,pc, it would have a peak flux density of only $S_{\nu} \sim$ 0.07\,mJy at 15\,MHz, which is too faint to be detectable with existing radio observatories. Nevertheless, there are more optimistic predictions for higher exoplanetary radio emission favouring planets with short (i.e., $<\,0.1\,$au) orbital distances (i.e., the so-called `hot Jupiters') orbiting young stars \citep[e.g.,][]{farrell_1999, zarka_2001, lazio_2004, stevens_2005, griessmeier_2007}. These studies have applied empirical scaling laws known to operate in our solar system to nearby exoplanetary systems and generally predict that a small number of hot Jupiter type exoplanets could have radio flux densities of the order of a few mJys and could be detectable with existing radio observatories. The preference in these models for hot Jupiters is akin to the reason why Earth, being closer to the Sun and having a larger incident solar wind power, is more luminous than Uranus or Neptune even though its magnetic field strength is much less. The same reasoning applies as to why young stars, having mass loss rates $1-2$ orders of magnitude greater than the solar value, are favoured in these models. Consequently, most detailed targeted searches for exoplanetary radio emission have focused on hot Jupiter type systems \citep[e.g.,][]{zarka_1997, bastian_2000, ryabov_2004, lazio_2007, lecavelier_2011, hallinan_2013}, but there has been no confirmed detections as of yet.

There are however, some inherent disadvantages in searching for radio emission from hot Jupiter type exoplanets. Tidal locking may reduce dynamo action and cause their internal magnetic fields to be very weak. For example, \cite{griessmeier_2004} predict that the magnetic moment of closely orbiting (i.e., $\leqslant\,0.05\,$au) Jupiter mass planets can be less than one tenth of the value observed for Jupiter. This would make such planets observable only at wavelengths below the Earth's ionospheric cut-off (i.e., $\nu < 10\,$MHz). We note however that \cite{reiners_2010} found that some hot Jupiters may possess sufficiently strong remnant magnetic fields so that radio emission could be produced above the Earth's ionospheric cut-off frequency, albeit rarely reaching 150\,MHz. \cite{zarka_2007} also discuss the possibility for hot Jupiters to excite radio emission above the Earth's ionospheric cut-off frequency from strongly magnetized stars (i.e., with stellar surface magnetic fields of the order $10^3\,$G). The radio emission from hot Jupiters is also likely to be modulated with the planetary orbital period \citep{hess_2011}. Therefore tidally locked hot Jupiters usually need to be observed for a few days to achieve full rotational phase coverage. For example, \cite{hallinan_2013} performed one of the most sensitive radio searches for any hot Jupiter system to date and yet only achieved 50\% rotational phase coverage in 40 hours of observations. We note that \cite{vasylieva_2015} did manage to observe a very short period ($\sim 20\,$hours) hot Jupiter for a total of 42\,hours therefore covering twice the orbital phase of this planet.

The inherent disadvantages posed by hot Jupiters can be circumvented by observing planets at larger orbital distances. \cite{nichols_2011} showed that rapidly rotating planets at large orbital distances (i.e., many au) that are subjected to high X-ray/UV illumination from their host star and have a plasma source within their magnetosphere (e.g., a volcanic exomoon) could produce detectable radio emission. In general however, longer period systems will be intrinsically fainter than hot Jupiters for a given stellar mass loss rate and so planets immersed in dense stellar winds are favourable when searching for emission from these systems. \cite{george_2007} used the Giant Metrewave Radio Telescope (GMRT) to observe two planets at relatively large orbital distances ($\geq 1\,$au) around young main sequence stars ($\epsilon$ Eri and HD\,128311) with stellar winds approximately 20 times more dense than the solar wind, but failed to detect any emission.

Evolved stars, that is stars that are post main sequence evolving, are a largely unexplored parameter space in the search for exoplanetary radio emission. Their huge mass loss rates can be many orders of magnitude greater than the values of young stars and so long period planets immersed in the dense winds of evolved stars are promising targets in the search for exoplanetary radio emission. \cite{ignace_2010} argued that long period planets around evolved stars with fully ionized winds - the so called `coronal giants' - may produce radio emission at the milli-Jansky level, i.e., levels of emission detectable with existing facilities. They also argued that planets around evolved stars with more neutral winds - winds from asymptotic-giant branch (AGB) stars and late spectral type red giants - are much weaker emitters and should not be detectable with existing facilities. \cite{fujii_2016} showed that the accretion process of a mainly neutral evolved star's wind onto a planet would emit UV and X-ray photons which would ionize the stellar wind in the planet's vicinity and enhance the radio signal. However, their predicted levels of radio emission from these systems would still not be detectable with existing facilities.

In this paper, we extend the search for exoplanetary radio emission to planets around evolved stars. We use the Low-Frequency Array (LOFAR) to carry out a deep pointed search of three coronal giant stars with known exoplanets to search for low frequency radio emission at 150\,MHz. Previous studies have shown these stars to be weak thermal emitters  at centimeter wavelengths with no non-thermal component \citep{ogorman_2016} and we can therefore expect the stellar emission to be completely negligible at 150\,MHz (i.e., $<\,1\,\mu$Jy). In Section \ref{sec2} we discuss the properties of our three targeted systems and explain our reasons for observing them. In Section \ref{sec3} we present our LOFAR observations and explain our data reduction strategy. Our results are presented in Section \ref{sec4} and a discussion of these is presented in Section \ref{sec5}. Finally, we summarize our findings in Section \ref{sec6}.

\section{Radio emission predictions and target selection}\label{sec2}

\begin{table*}
\caption{Basic parameters of the observed evolved stars and their planetary companions.}
\label{tab1}
\centering
\begin{tabular}{c c c c c c c c c c c c}
\hline\hline
\rule{0pt}{1.0\normalbaselineskip}
					&	&  & Host star  & &     & & & &  Planet & &  \\
\cmidrule{1-7} \cmidrule{9-12}
Source					& Spectral 			    & $d$            & R$_{\star}$      		& M$_{\star}$   & $\upsilon _{\infty}$ & $\dot{M}_{\textrm{ion}}$ &	& $\mathit{M}_{\textrm{P}}\textrm{sin}i$         & $a$    & $S_{\nu}$  \\
						& 			 type       			& (pc)		         &  (R$_{\odot}$)   & (M$_{\odot}$)    & (km s$^{-1}$) & (M$_{\odot}$ yr$^{-1}$)& & (M$_{J}$)      &  (au)         & (mJy)  \\
\hline
\rule{-2.6pt}{2.5ex} $\beta$ Gem & K0\,III & $10.4$ & $8.8^{^a}$ & $1.9^{^b}$ & 215 & $3.1\times 10^{-11}$  & & $2.9^{^c}$& $1.7^{^c}$ &  14.6 \\
					 $\iota$ Dra & K2\,III & $31.0$ & $12.9^{^d}$ & $1.8^{^d}$ & 173 & $5.8\times 10^{-11}$ & & $12.6^{^d}$ &$1.3^{^e}$ & 1.7 \\
					 $\beta$ UMi & K4\,III & $40.1$ & $42.1^{^f}$ & $1.4^{^g}$ & 30 & $2.7\times 10^{-10}$& &$6.1^{^g}$ &$1.4^{^g}$& 0.1 \\
\hline
\end{tabular}
      \vspace{-2mm}
     \tablefoot{Distances, $d$, are based on parallaxes from \cite{van_leeuwen_2007}. R$_{\star}$, M$_{\star}$, $\upsilon _{\infty}$, and $\dot{M}_{\textrm{ion}}$ are the host star's radius, mass, wind terminal velocity, and ionized mass loss rate, respectively. $\mathit{M}_{\textrm{P}}\textrm{sin}i$, $a$, and $S_{\nu}$ are the planet's minimum mass, semi-major orbital axis, and expected radio flux density. The superscript letters represent the following references: (a) \cite{nordgren_2001}; (b) \cite{hatzes_2012}; (c) \cite{reffert_2006}; (d) \cite{baines_2011}; (e) \cite{frink_2002}; (f) \cite{richichi_2005}; (g) \cite{lee_2014}. }
\end{table*}

To date, more than 100 exoplanets have been detected around evolved stars\footnote{Taken from the Extrasolar Planetary Encyclopedia \citep{schneider_2011}.}. All of these planets are in orbits exterior to $\sim$\,0.5\,au and are on average more massive than planets around main sequence stars \citep[e.g.,][]{jones_2014}. The majority of these planets are found around first ascent giants while to date no planets have been confirmed around the more evolved asymptotic giant branch (AGB) stars. In choosing suitable evolved star targets as potential sources of exoplanetary radio emission we use an empirical scaling relationship known as the ``radiometric Bode's law'' (RBL) \citep{desch_1984}. This law is based on observations of the magnetized solar system planets and relates the planet's median emitted radio power to the incident solar wind power deposited onto the planet's  magnetosphere. The energy source is believed to be either the kinetic energy from the solar wind or a magnetic energy flux. Numerous studies have already extrapolated this law to exoplanetary systems \citep[e.g.,][]{farrell_1999,zarka_2001,lazio_2004}. Considering the goal of this study is to examine the effect of the kinetic energy of evolved stellar winds on exoplanetary radio emission we only use the kinetic energy RBL here. We stress that a radio-to-magnetic scaling law has also been proposed by \cite{zarka_2001} and \cite{zarka_2007}. Therefore, the possible limitations of this study by solely considering the radio-to-kinetic scaling law are discussed in Section \ref{sec5.4}. In Appendix \ref{ap1} we derive a variant of the kinetic energy RBL which includes the effects of the larger mass-loss rates and slower wind velocities of evolved stars in comparison to solar type stars and discuss our differences with a similar attempt by \cite{ignace_2010}. We find the exoplanetary radio flux density to be
\begin{eqnarray}\label{eq1}
S_{\nu} &\approx& 4.6\,\textrm{mJy}\left(\frac{\omega}{\omega _J} \right)^{-0.2}\left(\frac{M_P}{M _J} \right)^{-0.33}\left(\frac{R_P}{R _J} \right)^{-3} \times \nonumber \\ && {} \left(\frac{\Omega}{1.6\,\textrm{sr}} \right)^{-1}\left(\frac{d}{10\,\textrm{pc}} \right)^{-2} \left(\frac{a}{1\,\textrm{au}} \right)^{-1.6}\times \nonumber \\ && {}  \left(\frac{\dot{M}_{\textrm{ion}}}{10^{-11}\,M_{\odot}\,\textrm{yr}^{-1}} \right)^{0.8} \left(\frac{\upsilon _{\infty}}{100\,\textrm{km}\,\textrm{s}^{-1}} \right)^{2} 
\end{eqnarray}
where $\omega$ is the rotation rate of the exoplanet (i.e., $2\pi /\omega $ is the planetary rotation period) and $\omega _{\mathrm{J}}$ is the rotation rate of Jupiter (i.e., $2\pi /\omega _{\mathrm{J}}= 10\,$hrs), $M_{\mathrm{P}}$ is the mass of the planet in Jupiter masses, $M_{\mathrm{J}}$, $R_{\mathrm{P}}$ is the radius of the exoplanet in Jupiter radii, $R_{\mathrm{J}}$, $\Omega$ is the beaming solid angle of the emission, $d$ is the Earth-star distance, $a$ is the semi-major axis of the planet's orbit in au, $\dot{M}_{\mathrm{ion}}$ is the stellar ionized mass-loss rate, and $\upsilon _\infty$ is the terminal velocity of the stellar wind. In deriving this expression we have followed \cite{farrell_1999} and assumed that the planet will emit ECM emission between the frequencies $0.5\nu _c$ and $\nu _c $, where $\nu _c $ is the maximum radiation frequency and is discussed further in Section \ref{sec5.2}.

Equation \ref{eq1} enables us to predict the exoplanetary radio flux density from nearby evolved stars with known exoplanets. In choosing our evolved star targets, an important wind property to consider is the ionization fraction because it is only the ionized component of the wind that couples with the planet's magnetosphere. The ``Linsky-Haisch dividing line'' in the giant branch of the H-R diagram near spectral type K1 and near spectral type G5 for the brighter giants, separates these stars based on their wind ionization properties \citep{linsky_1979}. Stars blueward of the dividing line possess fully ionized winds like the Sun, have mass-loss rates $\dot{M} < 10^{-10}$\,M$_{\odot}\,$yr$^{-1}$, and wind terminal velocities $\upsilon _{\infty} \sim 100$\,km\,s$^{-1}$, while stars on the redward side have winds with lower levels of ionization, larger mass-loss rates $\dot{M} < 10^{-8}$\,M$_{\odot}\,$yr$^{-1}$, and lower wind terminal velocities $\upsilon _{\infty} \leq 40$\,km\,s$^{-1}$ \citep{drake_1986}. In Table \ref{tab1} we list the basic parameters of two early-K type giants and one mid-K type giant that we observed with LOFAR. The two early-K giants, $\beta$ Gem and $\iota$ Dra, are weak x-ray emitters \citep{huensch_1996} which presumably originates from their thermal coronae, and so they most likely possess coronal winds that are fully ionized. The mid-K giant, $\beta$ UMi, may have a partially ionized wind with an ionization fraction of $\sim 0.2$, although this value is based on a low signal-to-noise centimeter observation of the star \citep{drake_1986}. The ionized mass-loss rates are calculated using the semi-empirical mass loss relation from \cite{schroder_2005} and applying an ionization fraction of 1 and 0.2 for the early- and mid-spectral type giants, respectively. To estimate the velocity of the fully ionized coronal type winds of the early-K type giants, we follow \cite{drake_1986} and assume $\upsilon _{\infty} = 0.75 \times v_{\textrm{esc}}$, where $v_{\textrm{esc}}$ is the photospheric escape velocity. The velocity of the partially ionized wind of the mid-K type giant is found from \ion{Mg}{II} absorption features \citep{drake_1986}. All three evolved stars have been confirmed to host at least one sub-stellar companion, whose semi-major axes are between 1.3 and 1.7\,au and have minimum masses greater than $2.9\,M_{\textrm{J}}$. We assume a beaming solid angle of 1.6 steradians for all three targets which is the same as that of Jupiter's decameter emission \citep{desch_1984, zarka_2004}. Following Equation \ref{eq1} we find that the predicted exoplanetary radio flux densities from the three evolved stars range from 0.1 to 14.6\,mJy. We note that these predictions are based on many system parameters that  are highly uncertain, particularly the ionized mass-loss rates which are poorly constrained by observations. \cite{ogorman_2016} used centimeter radio observations of $\beta$ Gem to place upper limits on the ionized mass-loss rate which were almost identical to the predictions of \cite{schroder_2005}. It is therefore likely that the ionized mass loss rates given in Table \ref{tab1} are upper limits to the actual values. The predicted exoplanetary radio flux densities in Table \ref{tab1} should be seen as a `zeroth order' estimate to show the feasibility of detecting such emission from evolved stars with LOFAR.

\section{LOFAR observations and data reduction}\label{sec3}
The three targets were observed with LOFAR \citep{van_haarlem_2013} over three nights between February and May 2015 using the high band antennas (HBA) (Program code: LC3-009, PI: Eamon O'Gorman). A brief overview of these observations is given in Table \ref{tab2}. Each target was observed over a single track lasting 8 hours in total with approximately 7 hours on source. Observations were taken with the entire LOFAR array (i.e., core, remote, and international stations) although data from only the core and remote stations are used in this paper. As it is requires greater effort to calibrate the data from the long baselines, we decided not to attempt to calibrate these, unless we detected emission at or close to the expected location of our targets in the data from the core and remote stations first. 

The observations were designed to allow for the calibration of all stations and a similar observational setup to that outlined in \cite{varenius_2015} was followed. For both $\beta$ Gem and $\iota$ Dra, two nearby calibrators were also observed simultaneously using a total of three beams, each covering a bandwidth of 31.64\,MHz centred at 150\,MHz using 162 sub-bands and 1 channel per sub-band (channel width 195\,kHz). The purpose of these two calibrators were to solve for the delay, rate and phase of the international stations. For $\beta$ UMi, the paucity of a nearby phase calibrator resulted in only one nearby delay and rate calibrator being observed simultaneously using a total of two beams, each covering a bandwidth of 47.66\,MHz centered at 150\,MHz using 244 sub-bands and 64 channels per sub-band (channel width 3\,kHz). The intent of the finer frequency sampling of the $\beta$ UMi data was to help locate a suitable nearby phase calibrator by reducing frequency averaging and thereby increasing the field of view (FOV). All observations were interleaved with a 2 minute observation of a flux calibrator every 30 minutes using a single beam which covered the same bandwidth as the target and nearby calibrators. The data were stored in the LOFAR long term archive (LTA) with integration times of 1.0 and  8.0\,s for the targets and calibrators, respectively. Finally, we used the LOFAR \textit{new default pre-processing pipeline} (NDPPP) to flag the data using AOFLAGGER \citep{offringa_2012}. NDPPP was also used to average the $\beta$ Gem data to a 5\,s integration time and the $\beta$ UMi data to a 5\,s integration time and 4 channels per subband (channel width 48.8\,kHz).

The PREFACTOR pipeline \citep{van_weeren_2016} was used to provide amplitude and clock calibration, as well as initial phase calibration, for all three target sources. PREFACTOR was used to calculate the amplitude gains for each individual flux calibrator scan using the default \cite{scaife_2012} calibrator models. Phase solutions were also derived and split into terms accounting for clock errors and atmospheric total electron content (TEC). Both the amplitude solutions and the clock portion of the phase solutions were then applied to the corresponding scan of the target field. The TEC solutions were not propagated as they were applicable only to the calibrator field. The diagnostic plots of the amplitude and phase solutions derived from the calibrator fields were examined for bad solutions, but there was no evidence that any scans needed to be excluded.

%
\begin{table*}
\caption{LOFAR 150\,MHz observations of the three evolved stars.}
\label{tab2}
\centering
\begin{tabular}{c c c c c c c c c c c}
\hline\hline
\rule{-3pt}{2ex}Target	& Date			& Bandwidth    & Time           & Flux       	& Delay/rate/	& Synthesized   & $\sigma _{\mathrm{rms}}$  \\
  				& 			&     & on source          & calibrator      & phase		& beam FWHM &  \\
				&   			& (MHz)	&(hr) 	         &     & calibrator(s) &($\arcsec\, \times\,\arcsec, ^{\circ}$)    &     ($\mu$Jy beam$^{-1}$)   \\
\hline
\rule{-3pt}{2ex}$\beta$ Gem & 2015 Feb 17,19 & 31.64& 7.0 & 3C196 & J0741+3112 &$7.5\times 3.8, 57$ &  325  \\
 &  & &  &  & J0746+2734 & &    \\
$\iota$ Dra & 2015 Apr 22,23 & 31.64 &6.5 & 3C295 & J1604+5714&$5.3\times 3.7, 91$&  290\\
 &  & &  &  & J1527+5849 & &    \\
$\beta$ UMi & 2015 May 05,06 & 47.66 & 7.0 & 3C295 &J1448+7601 & $7.3\times 5.4, 76$ &  190 \\
\hline
\end{tabular}
      \vspace{-2mm}
     \tablefoot{The delay, rate, and phase calibrators were included in our observational setup to enable the calibration of the international baselines if desired. Data from only the core and remote stations are used in this paper and so these calibrators were not used in our data reduction.}
\end{table*}

Following this initial amplitude and clock calibration, the data were combined into chunks of 10 LOFAR sub-bands (approximately 2~MHz) and direction-independent phase calibration was performed on the target field using an initial sky model taken from the LOFAR global sky model \citep[GSM;][]{van_haarlem_2013} and a combination of the VLA Low-frequency Sky Survey \citep[VLSS, VLSSr;][]{cohen_2007, lane_2012}, the Westerbork Northern Sky Survey \citep[WENSS;][]{rengelink_1997} and the NRAO VLA Sky Survey \citep[NVSS;][]{condon_1998}. A model of each field at LOFAR frequencies was then developed by imaging three 2~MHz chunks of the data at the beginning, middle and end of the total bandwidth and using the LOFAR source finding tool, PyBDSM \citep{mohan_2015}, to make a multi-frequency sky model of each field. The resulting models were then used as the basis for a further round of phase-only direction-independent self-calibration. This process was repeated until two rounds of self-calibration had been applied to the data. In the case of the $\iota$ Dra dataset, excessive averaging prior to running PREFACTOR meant that many sources away from the phase centre were badly distorted. The resulting sky models were not suitable for self-calibration and the output of the PREFACTOR script was used as the final calibrated dataset.

Each full dataset was then imaged with CASA using 2 Taylor terms to describe the variation of the flux with frequency for each source in the field being imaged. 256 w-projection plains were used to account for the curvature of the sky plane when modeling all strong sources in the FOV. Many sources in the FOV exhibited distortions due to beam or calibration issues. A suitable mask was generated using PyBDSM to exclude regions of probable spurious emission from the cleaning process at all times - from the calibration cycle to the final image. Data with a \textit{uv}-range of less than 1\,k$\lambda$ were excluded to reduce the presence of any diffuse emission detected by the shortest baselines. It should be noted that CASA does not implement any correction for the LOFAR beam, and while this option is available in AWImager \citep{tasse_2013}, AWImager lacks the option to use multiple Taylor terms to describe the sky brightness, leaving it at a disadvantage when imaging with a large fractional bandwidth. As each target source is located at the centre of the field, it was felt that primary beam effects were less important than correct treatment of the flux across the entire bandwidth. We note that the separation in elevation angles between $\beta$ Gem, $\iota$ Dra, and $\beta$ UMi and their flux density calibrators are about $10^\circ$, $5^\circ$, and $20^\circ$ respectively. Based on these separations, and the findings of \cite{coughlan_2017}, we estimate the absolute flux density uncertainty to be approximately 10\%. As well as imaging the entire dataset, separate images of each 30 minute scan were also generated with the same settings to search for any evidence of variable behavior or flaring in the target sources.

\section{Results}\label{sec4}
\begin{figure*}[bt!]
\includegraphics[clip,angle=90, scale=0.8]{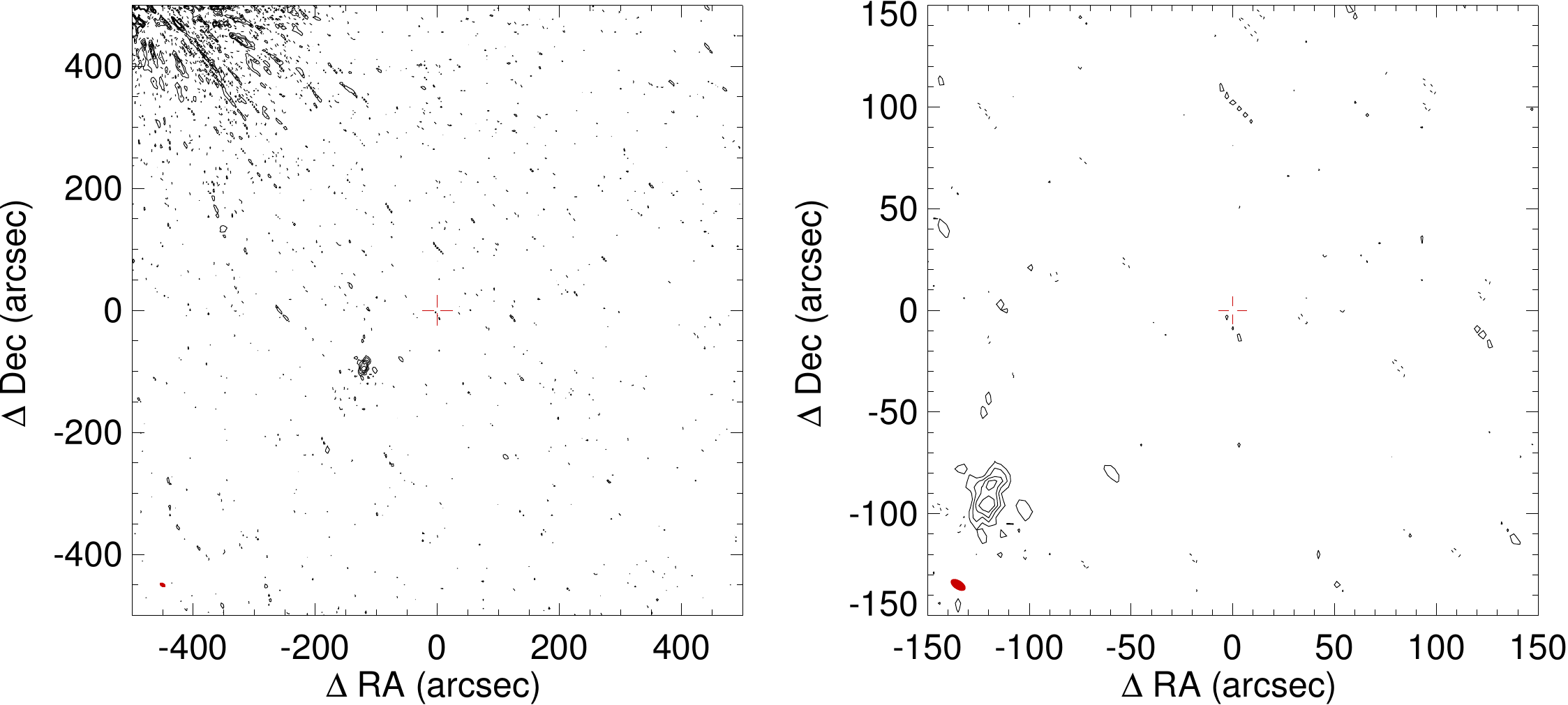}
\caption[]{\textit{Left panel:} LOFAR $1000\arcsec \times 1000\arcsec$ image of the field around $\beta$ Gem. \textit{Right panel:} A zoomed in $300\arcsec \times 300\arcsec$ version of the image shown in the left panel. The restoring beam FWHM is displayed in red in the lower left corner of both images and has dimensions $7.5\arcsec \times 3.8\arcsec$. Contours are set to $(-3,3,6,9,12,15,100,200,300)\times \sigma _{\mathrm{rms}}$, where $\sigma _{\mathrm{rms}} = 330\,\mu\,$Jy\,beam$^{-1}$ and is the rms noise at the target position. The red crosshairs mark the expected position of $\beta$ Gem at the epoch of observation.}
\label{fig0}
\end{figure*}

\begin{figure*}[bt!]
\includegraphics[clip,angle=90, scale=0.8]{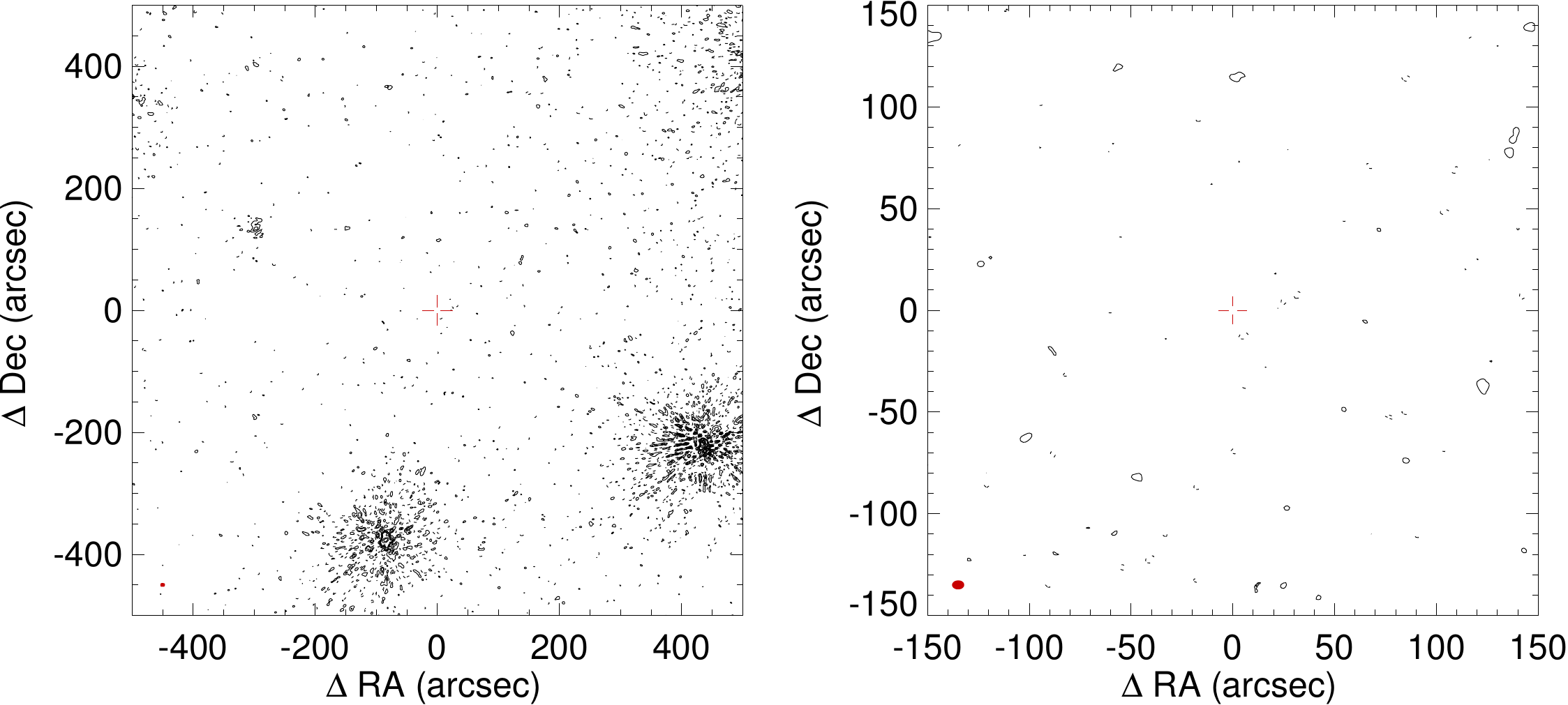}
\caption[]{\textit{Left panel:} LOFAR $1000\arcsec \times 1000\arcsec$ image of the field around $\iota$ Dra. \textit{Right panel:} A zoomed in $300\arcsec \times 300\arcsec$ version of the image shown in the left panel. The restoring beam FWHM is displayed in red in the lower left corner of both images and has dimensions $5.3\arcsec \times 3.7\arcsec$. Contours are set to $(-3,3,6,9,12,15,30,45,60)\times \sigma _{\mathrm{rms}}$, where $\sigma _{\mathrm{rms}} = 290\,\mu\,$Jy\,beam$^{-1}$ and is the rms noise at the target position. The red crosshairs mark the expected position of $\iota$ Dra at the epoch of observation.}
\label{fig1}
\end{figure*}

\begin{figure*}[hbt!]
\centering 
\includegraphics[clip,angle=90, scale=0.8]{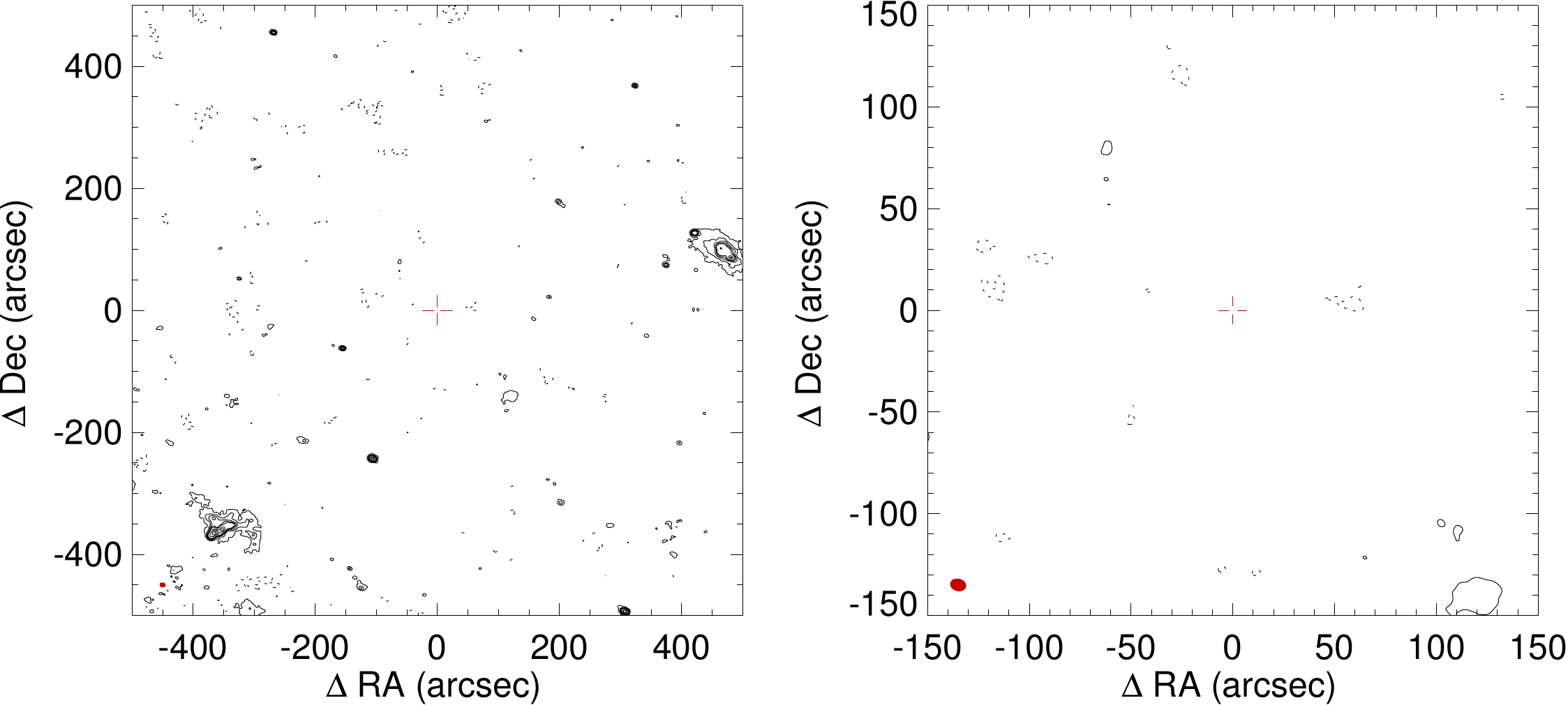}
\caption[]{\textit{Left panel:} LOFAR $1000\arcsec \times 1000\arcsec$ image of the field around $\beta$ UMi. \textit{Right panel:} A zoomed in $300\arcsec \times 300\arcsec$ version of the image shown in the left panel. The restoring beam FWHM is displayed in red in the lower left corner of both images and has dimensions $7.3\arcsec \times 5.4\arcsec$. Contours are set to $(-3,3,6,9,12,15,30,45)\times \sigma _{\mathrm{rms}}$, where $\sigma _{\mathrm{rms}} = 190\,\mu\,$Jy\,beam$^{-1}$ and is the rms noise at the target position. The red crosshairs mark the expected position of $\beta$ UMi at the epoch of observation.}
\label{fig2}
\end{figure*}

In Figures \ref{fig0}, \ref{fig1}, and \ref{fig2} we display both $1000\arcsec \times 1000\arcsec$ and $300\arcsec \times 300\arcsec$ images of the fields around each of the three targets at 150\,MHz. From these images, it can clearly be seen that we do not detect emission from any of the three evolved stars. The root-mean-square (rms) noise values, $\sigma _{\mathrm{rms}}$, at the expected location of each source, are found to be 0.325, 0.290, and 0.190\,mJy\,beam\,$^{-1}$, for the $\beta$ Gem, $\beta$ UMi, and $\iota$ Dra images, respectively. We can therefore place $3\sigma_{\mathrm{rms}}$ upper limits on the flux density at 150\,MHz of 0.98, 0.87, and 0.57\,mJy for $\beta$ Gem, $\beta$ UMi, and $\iota$ Dra, respectively. In the $\beta$ Gem image, we note that the rms noise increases to about 1\,mJy approximately one beam south-east of the expected source position which is caused by the sidelobes of a strong source located approximately $700\arcsec$ north-east of the target. 

To account for the fact that any exoplanetary radio emission may be time variable in nature, we also imaged each individual 30\,min scan for the three sources. Again, no emission was detected from the targets in these higher time resolution images, which typically had a sensitivity of $<1$\,mJy\,beam$^{-1}$ at the expected source position.
We did not attempt to construct a time series of the complex visibilities to search for even shorter time variable emission due to the very large FOV of LOFAR at 150\,MHz ($\sim\,6^{\circ}$) and the ensuing difficulty in modelling and removing all additional sources in the FOV.
\section{Discussion}\label{sec5}
\subsection{Results in the context of previous searches}\label{sec5.1}
There has been no confirmed direct detection of radio emission from an exoplanet to date despite a large number of dedicated searches. The first sensitive searches for radio emission from exoplanets were carried out with the Very Large Array (VLA). \cite{winglee_1986} observed six nearby main sequence stars of spectral type M with suspected substellar companions with the VLA at 1400 and 333\,MHz and achieved $3\sigma$ upper limits of 0.3 and 30\,mJy, respectively. \cite{bastian_2000} used the VLA to observe seven main sequence stars with known exoplanetary systems at 1465 and 333\,MHz, and one at 74\,MHz. Their typical $3\sigma$ upper limits were between $0.06 - 0.2\,$mJy (at 1465\,MHz), $3 - 30\,$mJy (at 333\,MHz), and $150\,$mJy (at 74\,MHz). \cite{lazio_2007} observed the hot Jupiter hosting main sequence F7V star $\tau$ Boo at 74\,MHz with the VLA over 3 epochs and reached $3\sigma$ upper limits of $\sim$300\,mJy for each epoch.

The 150\,MHz band of the Giant Metrewave Radio Telescope (GMRT) has also been used to search for exoplanetary radio emission. \cite{george_2007} reached $3\sigma$ upper limits of $\sim 14\,$mJy for exoplanetary emission from two young main sequence stars with the GMRT, while \cite{hallinan_2013} achieved an impressive $3\sigma$ upper limit of only $\sim 1.2\,$mJy after observing $\tau$ Boo for 40\,hours with the GMRT. \cite{sirothia_2014} surveyed 175 confirmed exoplanetary systems with the GMRT and reached a median $3\sigma$ upper limit value of $\sim 25\,$mJy for emission. They detected 4 radio sources coinciding with or located very close to known exoplanets but were unable to discriminate between the possibilities of background radio-sources and exoplanetary emission. \cite{lecavelier_2011, lecavelier_2013} also used the GMRT to observe two hot Jupiters and one hot Neptune around main sequence type stars and reported weak (i.e., $3\sigma \sim 3-4\,$mJy) emission close (within one synthesized beam) to two of their targets. However, their emission peaks appear to be consistent with noise peaks in their images. 

More recently, the Murchison Widefield Array (MWA) has also been used at 150\,MHz to search for exoplanetary radio emission. \cite{murphy_2015} targeted 17 known exoplanetary systems and placed $3\sigma$ upper limits in the range $15.2 - 112.5\,$mJy on the emission. \cite{lynch_2017} targeted exoplanets orbiting Myr-old stars in a region of recent star formation with MWA and achieved $3\sigma$ flux density limits down to 4\,mJy for highly polarized emission. 

Our LOFAR observations are the most sensitive to date in the search for exoplanets at 150\,MHz (i.e., at meter wavelengths). Our observations show that we can now routinely place $3\sigma$ upper limits at the sub-mJy level when searching for exoplanetary radio emission with LOFAR. They thus highlight the increased sensitivity that LOFAR now provides over other facilities at similar wavelengths such as the VLA, the GMRT, and the MWA. Moreover, the long baselines of the LOFAR remote stations provide superior angular resolution in comparison to the other aforementioned facilities. For example, our angular resolution (excluding the international baselines) is about $5\arcsec$ which is about 25 and 4 times better than that provided by the MWA and the GMRT at 150\,MHz, respectively, which helps to distinguish between any possible background confusing sources when imaging in Stokes \textit{I}. Furthermore, the International LOFAR baselines offer, if calibrated and imaged, a synthesized resolution of $0.25\arcsec$ at 150\,MHz (i.e., 80 times better than GMRT) which would be very useful to spatially constrain any detected candidate exoplanet emission.

\subsection{The importance of the observing frequency}\label{sec5.2}
Like the magnetized solar system planets, the ECM emission frequency of an an exoplanet is expected to occur at the electron gyrofrequency, $\nu (\textrm{MHz})= 2.8B\,(\textrm{Gauss})$, where $B$ is the planet's magnetic field strength. We note that $B$ can also be the host star's magnetic field in which an exoplanet may induce accelerated electron precipitation and radio emission in a similar fashion to the Io-Jupiter current system \citep{zarka_2007}. In the former case, there is also expected to be a sharp cut-off in this emission frequency at $\nu _c $, which will be controlled by the maximum planetary magnetic field strength, generally located close to the planet's surface at the magnetic poles. The maximum magnetic field strength for Jupiter is $\sim 14\,$G and so Jupiter does not emit ECM emission above 40\,MHz. Therefore, it is worth stressing that our 150\,MHz observations could never have detected radio emission from an exoplanet with a magnetic field strength similar to that of Jupiter's.  In fact, the exoplanets in our sample of evolved stars would need to generate magnetic field strengths of $\sim 50\,$G for us to have had a chance of detecting them.

A possible explanation for the lack of a detection of radio emission from exoplanets to date is that the majority of exoplanets observed do not generate sufficiently large magnetic field strengths to produce ECM emission at the observing frequency. This could be caused by a tendency to date for observing campaigns to focus on hot Jupiter type planets which, due to the $a^{-1.6}$ dependency in Equation \ref{eq1}, are predicted to produce greater levels of emission than planets with larger orbital radii. However, these planets are predicted to synchronize their axial spin with their orbital motion rapidly over their lifetime becoming tidally locked to their host star and so will be rotating much slower than Jupiter \citep{seager_2002}. Slowly rotating planets may have reduced dynamo efficiency and weaker  magnetic fields which would lower the ECM emission frequency \citep{griessmeier_2007}. For example, \cite{farrell_1999} showed that the maximum ECM emission frequency of a Jupiter size exoplanet can be approximated by
\begin{eqnarray}\label{eq2}
\nu_{c} \sim 23.5\,\textrm{MHz}\left(\frac{\omega}{\omega _J} \right)\left(\frac{M _P}{M _J} \right)^{5/3}.
\end{eqnarray}
Using the parameters of the well known hot Jupiter, $\tau$ Boo b ($2\pi/\omega  = 79\,$hrs, $M_P = 3.9\,M_J$) we find that $\nu _c = 28\,$MHz which would not be detectable with LOFAR at 150\,MHz. 

We reiterate that because the orbital distances of our observed planets are $>1.4\,$au, they are unlikely to be tidally locked and their rotation rates should be larger than hot Jupiter type planets. Assuming the planets in our sample have Jupiter rotation rates and masses that are equal to their minimum mass (i.e., $\textrm{sin}i=1$), then the maximum emission frequencies of our sample should have a range between $140-1600\,$MHz. These values could be larger or smaller by a factor of about 3 \citep{lazio_2004}. Along with the uncertainties in planetary masses and rotation rates, it is reasonable to suggest that the planets in our sample could be capable of producing ECM emission at 150\,MHz.

\subsection{Time variable emission}\label{sec5.3}
The low frequency radio emission from the magnetized solar system planets is highly variable over time. For example, the non-Io controlled decameter emission of Jupiter varies smoothly over minutes and can produce flux densities that are 10 times higher than the median levels \citep{zarka_1998,zarka_2004, marques_2017}, while Saturn's low frequency radio emission can be 100 times higher over similar durations \citep{desch_1985}. The causes of this variability can be due to modulation from the planetary spin rotation (due to the misalignment between the spin axis and the magnetic axis) and/or variability of the solar wind. Moreover, decameter emission of Jupiter is only detectable over certain ranges of rotational phase because the emission is narrowly beamed (see \citealt{hallinan_2013} for a discussion). It is therefore conceivable that any variable emission could have been diluted and have gone undetected due to our relatively long integration times ($\sim\,7\,$hrs). For example, from Equation \ref{eq1} we find that the expected median level of exoplanetary radio flux from $\beta$ Gem to be 14.3\,mJy while we reached a $3\sigma$ noise level of 0.325\,mJy over an integration time of 7 hours. If the planet produced only one burst of emission of 14.3\,mJy lasting say 5\,min then the flux would be diluted by 14.3\,mJy(5\,min/420\,min) $=$ $0.17\,$mJy and we would not have detected this emission (at a $3\sigma$ significance level). Similarly, we would not have detected this emission in our 30\,min integration images because the statistical noise in these were $\sim\,1\,$mJy and the flux would have been diluted to $\sim\,2\,$mJy. However, as the integration time, $t$, reduces we can roughly assume that the sensitivity reduces as $t^{-1/2}$, and so an image created over only 5\,min would have had a noise level of $\sim\,3\,$mJy and we could have detected the emission burst. We have found that the poor $u-v$ coverage makes imaging on these short (i.e., minutes) timescales difficult. Our observing strategy is therefore more suited to exoplanetarary radio emission that either varies smoothly over hours or is continuously bursty over hours. One future possibility to compensate for the poor $u-v$ coverage over these short time periods is to  observe the target over the entire 96\,MHz bandwidth and use the Multi-Frequency Synthesis (MFS) algorithm \citep{rau_2011} when imaging.

\subsection{Possible limitations of our model}\label{sec5.4}
\cite{zarka_2001} proposed a radio-to-magnetic scaling law whereby the conversion of incident solar wind magnetic energy into electron acceleration could be responsible for planetary radio emission in our solar system. They showed that the kinetic and magnetic energy flux of the solar wind vary similarly with distance beyond $\sim 1\,$au, and so from the two observed scaling laws alone, it is not possible to tell which physical interaction actually drives planetary radio emission. Investigations of the radio emission from satellite-Jupiter interactions actually suggest that the physically grounded scaling law is the radio-to-magnetic one, while the radio-to-kinetic one may be just a coincidence \citep{zarka_2007}. Subsequent papers further discuss the physical mechanisms governing the radio-to-magnetic scaling law \citep[e.g.,][]{saur_2013}. It is therefore possible that our LOFAR non-detections indicate that the radio-to-kinetic scaling law is an invalid assumption.

As evolved stars expand and spin down, their surface magnetic field strengths are expected to be weaker than those of main sequence stars \citep{simon_1989}. Nevertheless, there is now a growing amount of evidence that evolved stars across a wide range of spectral types are magnetically active \cite[e.g.,][]{van_doorsselaere_2017,karovska_2005, lebre_2014, vlemmings_2005}. Indeed, a surface-averaged longitudinal magnetic field of 0.5\,G has been measured for one of our evolved star targets, $\beta$ Gem  \citep{auriere_2009}, which is approximately a factor of three smaller than the solar value. In this case, given the larger stellar radius of $\beta$ Gem ($R_{\star} = 8.8\,R_{\odot}$), its magnetic moment would be larger than the solar value which could also increase the exoplanetary radio emission. It might be possible that magnetically active late spectral type evolved stars, whose stellar radii are much larger than our three targets, could have very large magnetic moments and thus be favourable candidates for exoplanetary radio emission.

\section{Conclusions}\label{sec6}
We have derived a variant of the radiometic Bode's law which accounts for the different stellar wind properties of evolved stars in comparison to solar type stars and used this to make an order of magnitude estimate for the expected levels of exoplanetary radio emission around a small sample of evolved stars. Our findings are in agreement with \cite{ignace_2010} in that some evolved stars with very ionized winds should be good targets to search for exoplanetary radio emission and may produce emission that is detectable with current radio facilities, provided the radio-to-kinetic scaling law is a valid assumption. We used LOFAR at 150\,MHz to search for exoplanetary radio emission from three such evolved stars, $\beta$ Gem, $\iota$ Dra, and $\beta$ UMi, all of which have known planetary companions. We did not detect any of these sources but place tight $3\sigma$ upper limits of 0.98, 0.87, and $0.57\,$mJy on their flux densities, assuming non-variable emission. 

There are good reasons to continue the search for exoplanetary radio emission from nearby evolved stars with relatively ionized winds. Not only might their large ionized mass loss rates enable them to produce levels of emission that are detectable with existing radio facilities, but importantly, planets at large orbital distances could be detectable from these objects. This surmounts a major obstacle when observing hot Jupiter type planets, namely that the planet will not be tidally locked to its host star and has a good chance of producing ECM emission at the frequencies observed.  

There are a host of possible reasons to explain why we and others have so far failed to detect radio emission from exoplanets (see \citealt{bastian_2000} and \citealt{zarka_2015}). An obvious strategy for increasing the likelihood of a detection is to observe as large a sample as possible with the most sensitive long wavelength radio telescopes available. LOFAR is currently the best instrument for this task and is easily capable of reaching $3\sigma$ upper limits at the sub-mJy level over typical observing periods of a few hours, which was not possible with previous long wavelength radio telescopes. LOFAR thus provides us with the best chance yet to detect this elusive emission. In the near future, the Square Kilometre Array (SKA) is expected to be over an order of magnitude more sensitive than LOFAR and is expected to operate at frequencies above 50\,MHz making it an ideal instrument for searching for radio emission from exoplanets \citep{zarka_2015}. It should be easily capable of detecting radio emission from nearby massive exoplanets that are not tidally locked to their host stars.

\begin{acknowledgements}
LOFAR, the Low Frequency Array designed and constructed by ASTRON, has facilities in several countries, that are owned by various parties (each with their own funding sources), and that are collectively operated by the International LOFAR Telescope (ILT) foundation under a joint scientific policy. The project was supported by funding from the European Research Council under the European Union’s Seventh Framework Programme (FP/2007-2013) / ERC Grant Agreement n. 614264 and a grant from the Irish Research Council. The authors wish to acknowledge the DJEI/DES/SFI/HEA Irish Centre for High-End Computing (ICHEC) for the provision of computational facilities and support.
\end{acknowledgements}


\bibliographystyle{aa}
\bibliography{references}
\begin{appendix} 
\section{Derivation of RBL for evolved stars}\label{ap1}
To estimate the predicted median radio flux density from an exoplanet around an evolved star we first follow \cite{lazio_2004} and use the `Radiation Model 2' from \cite{farrell_1999} which itself is based on the solar system work of \cite{desch_1984}. This model assumes that the emitted radio power from the exoplanet, $P_{\textrm{rad}}$, is related to the incident kinetic power from the stellar wind onto the magnetosphere,  $P_{\textrm{sw}}$, such that  $P_{\textrm{rad}} \propto  P_{\textrm{sw}}^{1.2}$. The radius of the exoplanet's magnetosphere, $R_{\textrm{M}}$, orbiting an evolved star will depend on the stellar wind velocity, $\upsilon _{\infty}$, and the ionized mass loss rate of the host star, $\dot{M}_\textrm{ion}$, and is determined from a pressure balance between the exoplanet's magnetic field pressure and the stellar wind dynamic pressure giving
\begin{eqnarray}\label{eqa1}
R_{\textrm{M}} \propto (aM_B)^{1/3}(\dot{M}_\textrm{ion}\upsilon _{\infty})^{-1/6}
\end{eqnarray}
where $a$ is the semi-major axis of the planet's orbit and $M_B$ is the planetary magnetic moment \citep[e.g.,][]{griessmeier_2005}. Following \cite{farrell_1999} a Blackett type scaling law relates the magnetic moment to the planet's mass, $M_p$, and rotation rate, $\omega$ ($\omega = 2\pi /$rotation period), such that $M_B \propto \omega M_p^{5/3}$, and so  
\begin{eqnarray}\label{eqa2}
R_{\textrm{M}} \propto (a\omega)^{1/3}M_p^{5/9}(\dot{M}_\textrm{ion}\upsilon _{\infty})^{-1/6}.
\end{eqnarray}
Assuming a spherically symmetric wind with  an ionized density of $\rho = \dot{M}_\textrm{ion}/4\pi a^2\upsilon _{\infty}$, the stellar wind power onto the magnetosphere can be written as 
\begin{eqnarray}\label{eqa3}
P_{\textrm{sw}} \propto \dot{M}_\textrm{ion}\upsilon _{\infty}^{2}R_{\textrm{M}}^{2}a^{-2}
\end{eqnarray}
where we have utilized Equation 1 of \cite{farrell_1999}. The emitted radio power from the exoplanet is then
\begin{eqnarray}\label{eqa4}
P_{\textrm{rad}} \propto \dot{M}_\textrm{ion}^{0.8}\upsilon _{\infty}^{2}a^{-1.6}\omega^{0.8}M_p^{1.33}.
\end{eqnarray}

Following \cite{farrell_1999}, the Jovian decametric component is considered at least partly related to the solar wind kinetic energy input and is used as the base power level to give
\begin{eqnarray}\label{eq5}
P_{\textrm{rad}} &=& 4\times 10^{18} \, \textrm{erg\,s}^{-1} \left( \frac{\dot{M}_\textrm{ion}}{10^{-14}\,M_{\odot}\,\textrm{yr}^{-1}} \right)^{0.8} \left( \frac{\upsilon _{\infty}}{400\,\textrm{km}\,\textrm{s}^{-1}}\right)^{2} \nonumber \\ && {} \left(\frac{a}{5\,\textrm{au}}\right)^{-1.6}\left( \frac{\omega}{\omega _{J}}\right)^{0.8}\left(\frac{M_p}{M_J}\right)^{1.33}
\end{eqnarray}
where the `J' subscripts indicate values for Jupiter and we assume that the solar mass loss rate is $\dot{M}_{\odot} = 10^{-14}\,M_{\odot}\,\textrm{yr}^{-1}$ and the solar wind velocity is $v_{\odot} = 400\,\textrm{km}\,\textrm{s}^{-1}$ at Jupiter's orbital distance of 5\,au.

The predicted radio flux density of an exoplanet is then 
\begin{eqnarray}\label{eq6}
S_{\nu} = \frac{P_{\textrm{rad}}}{\Delta \nu \Omega d^2}
\end{eqnarray}
where $d$ is the Earth-star distance, $\Omega$ is the beaming solid angle ($4\pi$\,sr being an isotropic pattern outwards), and $\Delta \nu$ is the emission bandwidth which is assumed to be $0.5\nu _c$ and is consistent with the solar system planets \citep{farrell_1999}. From Equation 4 of \cite{lazio_2004} the characteristic emission frequency is 
\begin{eqnarray}\label{eq7}
\nu _c \approx 23.5 \,\textrm{MHz} \left( \frac{\omega}{\omega _J}\right) \left( \frac{M_p}{M_J}\right)^{5/3} \left(\frac{R_P}{R_J} \right)^3
\end{eqnarray}
where $R_p$ is the planetary radius. Substituting Equations \ref{eq5} and \ref{eq7} into Equation \ref{eq6} gives
\begin{eqnarray}\label{eq8}
S_{\nu} &\approx& 4.6\,\textrm{mJy}\left(\frac{\omega}{\omega _J} \right)^{-0.2}\left(\frac{M_P}{M _J} \right)^{-0.33}\left(\frac{R_P}{R _J} \right)^{-3} \times \nonumber \\ && {} \left(\frac{\Omega}{1.6\,\textrm{sr}} \right)^{-1}\left(\frac{d}{10\,\textrm{pc}} \right)^{-2} \left(\frac{a}{1\,\textrm{au}} \right)^{-1.6}\times \nonumber \\ && {}  \left(\frac{\dot{M}_{\textrm{ion}}}{10^{-11}\,M_{\odot}\,\textrm{yr}^{-1}} \right)^{0.8} \left(\frac{\upsilon _{\infty}}{100\,\textrm{km}\,\textrm{s}^{-1}} \right)^{2} 
\end{eqnarray}

We note that \cite{ignace_2010} also applied the radiometric Bode's law to evolved stars but their equation differs from Equation \ref{eq8} in the following three ways: (1) The exponent of their solid angle term is $-2$ but should be $-1$. (2) Their formula contains a frequency term which should not be included. (3) They do not include the 1.2 exponent in the last term of their Equation 3 (i.e., the ratio of density times velocity cubed term) nor do they account for the variation in magnetospheric radius due to variations in stellar wind properties and so the exponent of their ionized mass loss rate is different to ours. Nevertheless, their basic conclusion is the same as ours in that exoplanets around coronal giants  could be good targets to search for exoplanetary radio emission.
\end{appendix} 
\end{document}